\documentclass[10pt,conference]{IEEEtran}
\IEEEoverridecommandlockouts
\usepackage[utf8]{inputenc}
\usepackage{graphicx} 
\usepackage{booktabs} 
\usepackage{amsmath}
\usepackage{bm}
\usepackage{gensymb}
\usepackage{physics}
\usepackage{mathtools, nccmath}
\usepackage{algorithm}
\usepackage{algpseudocode}
\usepackage{setspace}
\usepackage{amsfonts}
\usepackage{amssymb}
\usepackage{bbm}
\usepackage{cite}
\usepackage{lipsum}
\usepackage{authblk}
\usepackage{adjustbox}

\title{ODMA-Based Cell-Free Unsourced Random Access with Successive Interference Cancellation}

\author[*]{Mert Ozates}
\author[**]{Mohammad Kazemi}
\author[***]{Eduard Jorswieck}
\author[**]{Deniz Gündüz}
\affil[*]{IHP - Leibniz Institute for High Performance Microelectronics, 15236 Frankfurt (Oder), Germany \protect\\ Email: oezates@ihp-microelectronics.com}
\affil[**]{Department of Electrical and Electronic Engineering, Imperial College London  \protect\\ Email:\{mohammad.kazemi, d.gunduz\}@imperial.ac.uk}
\affil[***]{Institute for Communications Technology, Technische Universität Braunschweig, Germany \protect\\Email:  e.jorswieck@tu-braunschweig.de}

\date{}

\begin{document}

\maketitle

\begin{abstract}
We consider the unsourced random access problem with multiple receivers and propose a cell-free type solution for that. In our proposed scheme, the active users transmit their signals to the access points (APs) distributed in a geographical area and connected to a central processing unit (CPU). The transmitted signals are composed of a pilot and polar codeword, where the polar codeword bits occupy a small fraction of the data part of the transmission frame. The receiver operations of pilot detection and channel and symbol estimation take place at the APs, while the actual message bits are detected at the CPU by combining the symbol estimates from the APs forwarded over the fronthaul. The effect of the successfully decoded messages is then subtracted at the APs. Numerical examples illustrate that the proposed scheme can support up to 1400 users with a high energy efficiency, and the distributed structure decreases the error probability by more than two orders of magnitude.
\end{abstract}



\section{Introduction}

Unsourced random access (URA) \cite{polyanskiy} is an emerging paradigm addressing the dense traffic in massive machine-type communications (mMTC) applications, which is one of the main aspects of 5G and beyond communication systems. It is proposed for uplink communication of a massive number of users with small payloads. The communication is sporadic, i.e., only a small subset of the users are active at any given time and uncoordinated. This alleviates the problem of delay and signaling overhead that would be prohibitive with a huge number of users. In URA, all devices share the same codebook; hence, the user identities are eliminated, and the receiver aims to recover only a list of the transmitted messages.




In \cite{polyanskiy}, an achievability bound characterizing the core performance of URA over the Gaussian multiple access channel (MAC) is developed, and several low-complexity coding solutions based on slotted transmissions \cite{vem}, coded compressed sensing \cite{comp2}, random spreading \cite{ javad}, and on-off division multiple access (ODMA) \cite{odma2} are proposed. Also, the more practical fading MAC with URA is studied over flat \cite{nassaji} and frequency-selective fading \cite{ozates1}.

Massive multiple-input multiple-output (MIMO) \cite{massmimo} is a mature technology in 5G that provides high data rates and spatial multiplexing gains due to the significant degrees of freedom. It is also utilized in URA to increase the supported number of active users and energy efficiency \cite{fasura2}-\hspace{-0.1mm}\cite{odma4}. However, in massive cellular MIMO systems, the users at the cell edges can suffer from severe path loss. To cope with this issue, a new network paradigm called cell-free massive MIMO \cite{cellf1,cellf2} is proposed. In cell-free massive MIMO, each device in the network is served by many access points (APs) simultaneously. The APs are distributed in the same geographic area and connected to a central processing unit (CPU) via a fronthaul. In this way, the cell boundaries are removed, leading to lower path losses and improved spectral efficiency. However, since all APs serve all users, the original form of cell-free massive MIMO is not scalable; namely, the system complexity is unbounded with an unbounded number of users. This problem is solved in \cite{cellf2} by user-centric clustering, where each AP serves a subset of the users, which keeps the complexity finite even if the number of users is unbounded.



Although it is a suitable technology to support many users, the cell-free structure is examined in only a few works in the context of URA \cite{cefura,cellfura2,cellfura3}. A scalable cell-free URA scheme is proposed in \cite{cefura} where the users transmit a pilot signal and utilize the random spreading approach in the data part similar to \cite{fasura2}, while \textit{Level 2} cooperation \cite{cellf1} is employed to combine the symbol estimates. On the other hand, the received signals from different APs are jointly processed using variations of approximate message passing (AMP) in \cite{cellfura2} and \cite{cellfura3}. In the former, a location-based partitioned URA codebook matrix is utilized, while the latter employs the combination of channel coding and fixed sparse coding for transmission.


In this paper, we study a cell-free environment and propose a URA scheme utilizing the ODMA approach. Assuming that the transmission frame is divided into the pilot and data parts, each user first transmits a pilot sequence determined by a part of its message bits. The rest of the bits are encoded by a polar code and modulated prior to being distributed into the data part based on an on-off pattern. To have a scalable system \cite{cellf2}, only a subset of the active users are recovered at each AP, where the active pilot sequences and channel vectors are jointly estimated by the orthogonal matching pursuit (OMP) algorithm followed by linear minimum mean square error (LMMSE) symbol estimation. The symbol estimates are passed to the CPU, where an average of them is computed, namely, \textit{Level 2} cooperation among the APs is assumed. Then, a list decoder is employed to recover the transmitted message bits, and the successfully recovered messages are passed back to the APs for successive interference cancellation (SIC). Numerical examples demonstrate that the performance of the proposed scheme is superior to that of \cite{cefura} by up to 4.5 dB, and it can increase the number of supported active users up to 1400. Furthermore, distributing the receive antennas through the APs can decrease the system per-user probability of error (PUPE) by more than two orders of magnitude compared to the centralized structure.


The rest of the paper is organized as follows. We present the system model in Section \ref{system} and the proposed scheme in Section \ref{proposed}. We give a set of numerical results in Section \ref{results}, and conclude the paper in Section \ref{conclusion}.


\section{System Model} \label{system}

\begin{figure}
    \centering
    \hspace*{15mm}
     \includegraphics[scale = 0.45]{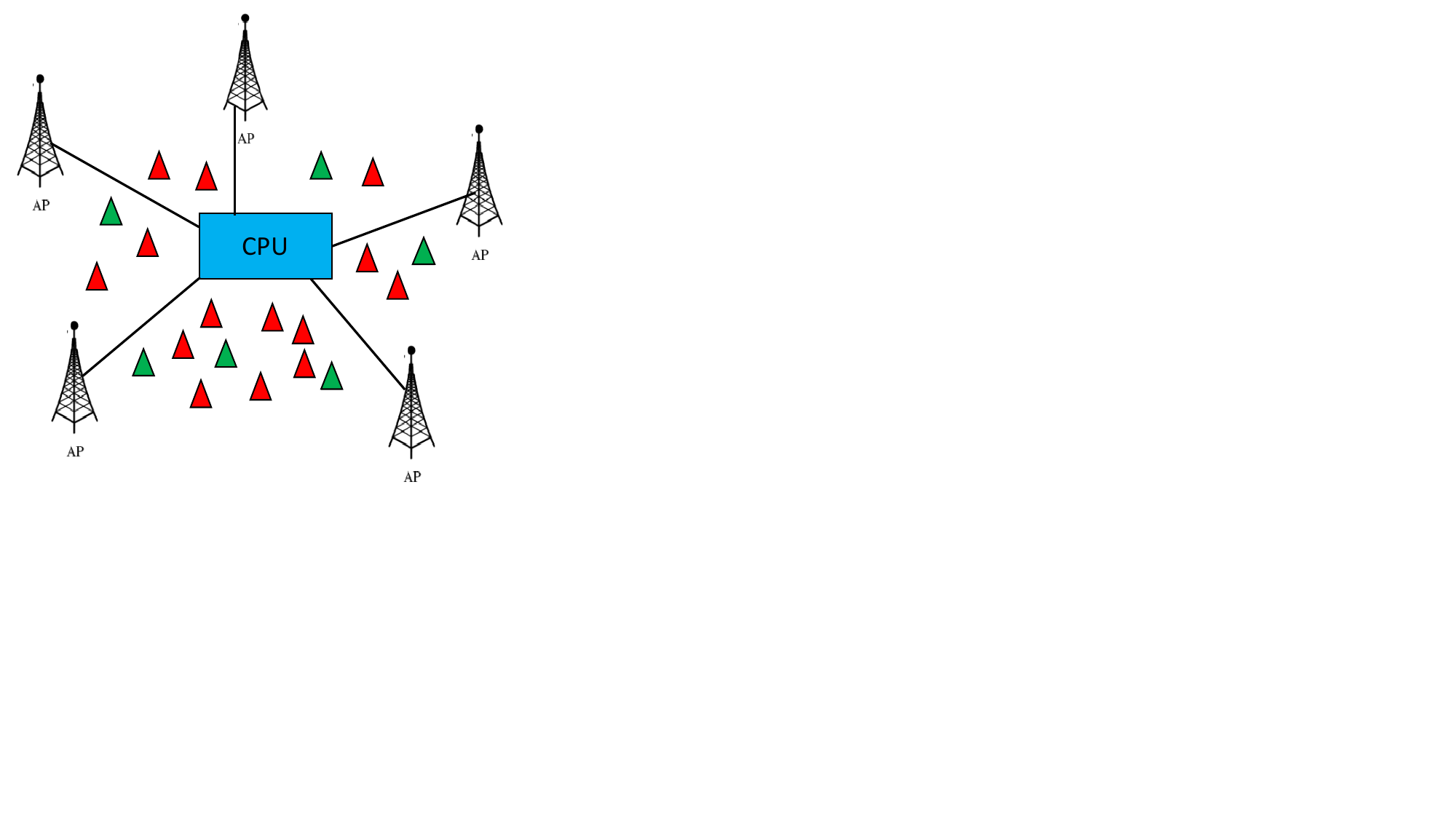}
     \vspace*{-42mm}
    \caption{Illustration of a cell-free system with 5 APs and 1 CPU. Active and inactive users are shown by green and red triangles, respectively.}
    \label{figsystem}
\end{figure}

We consider a cell-free scenario with a massive number of users equipped with a single antenna, while only $K_a$ out of $K_{\text{tot}}$ ($K_a \ll K_{\text{tot}}$) users are active in a given time interval. The active users transmit $B$ bits to $M$ APs without any coordination through a transmission frame of length $n$. The APs are equipped with $M_r$ antennas and randomly located in a $D \times D$ $m^2$ area. An example of such a system with 5 APs connected to a CPU is shown in Fig. \ref{figsystem}. The received signal at the $m$-th AP can be written as


\begin{equation}
    \mathbf{Y}_m = \sum\limits_{i =1 }^{K_a} \mathbf{x}_i \mathbf{g}_{i,m} + \mathbf{Z}_m,
\end{equation}

\noindent where $\mathbf{Y}_m \in\mathbb{C}^{n \times M_r}$, $\mathbf{x}_i \in\mathbb{C}^{n \times 1} $ is the transmitted signal of the $i$-th user that is obtained by encoding and modulation of the message of the $i$-th user $\mathbf{m}_i$, $\mathbf{g}_{i,m} \in\mathbb{C}^{1 \times M_r}$ is the vector of channel coefficients of the $i$-th user at the $m$-th AP, and $\mathbf{Z}_m \in\mathbb{C}^{n \times M_r}$ is the circularly symmetric additive white Gaussian noise (AWGN) with i.i.d. elements, each of them follows the distribution $\mathcal{CN} (0,\sigma^2)$, where $\mathcal{CN}$ denotes the complex normal distribution and $\sigma^2$ is the noise variance. The channel vector $\mathbf{g}_{i,m}$ is defined as \cite{cefura}

\begin{equation}
    \mathbf{g}_{i,m} = \sqrt{\beta_{i,m}} \mathbf{h}_{i.m},
\end{equation}

\noindent where $\mathbf{h}_{i.m}$ is the small-scale fading channel vector and ${\beta_{i,m}}$ is the large-scale fading coefficient of the $i$-th user at the $m$-th AP. We consider a quasi-static fading model, i.e., the fading coefficients remain constant throughout the transmission frame.





The APs are connected to a CPU via a fronthaul, where the aim is to produce a list of the transmitted messages using the symbol estimates from the APs. The system performance is measured in terms of the PUPE $P_e$ that can be calculated as $P_e = P_{\text{md}} + P_{\text{fa}}$, where $P_{\text{md}}$ and $P_{\text{fa}}$ are the misdetection and false alarm probabilities, which can be written as 

\begin{equation}
  P_{\text{md}} = \frac{1}{K_a} {\mathbb{E}\bigg[\sum\limits_{i \in \mathcal{K}_a} \mathbbm{1}_{\{\mathbf{m}_i \notin \mathcal{L}\}}\bigg]},
\end{equation}

\begin{equation}
   P_{\text{fa}} = \mathbb{E} \left[\frac{\abs{\mathcal{L} \setminus  \{\mathbf{m}_i: i \in \mathcal{K}_a \}}}{\abs{\mathcal{L}}}\right],
\end{equation}

\noindent where $\mathcal{K}_a$ is the set of active users, $\mathbbm{1}_{\{\ \hspace{-1mm} \cdot \}}$ is the indicator function and $\abs{\cdot}$ denotes the cardinality of a set. 

\section{Proposed Scheme} \label{proposed}

\subsection{Encoding}

\begin{figure*}
    \centering
     \includegraphics[scale = 0.42]{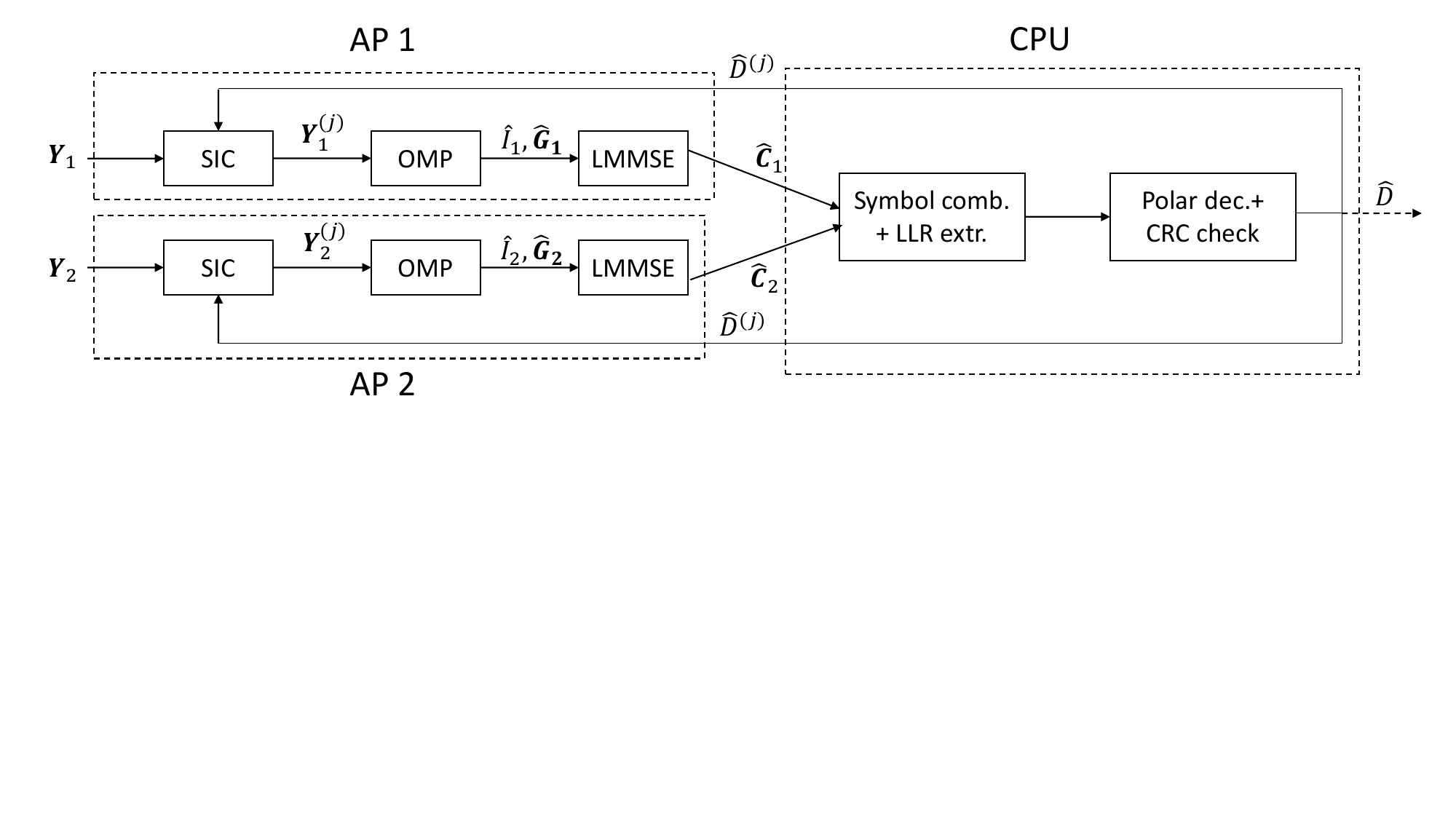}
     \vspace*{-43mm}
    \caption{An illustration of the decoding process of the proposed scheme with 2 APs.}
    \label{figdecoder}
\end{figure*}

At the user equipment side, we assume that each user maps its first $B_p$ bits to a pilot sequence of length $n_p$ that is selected from a common non-orthogonal pilot codebook $\mathbf{A} \in\mathbb{C}^{n_p \times N}$ with $N = 2^{B_p}$ candidate pilot sequences. The selected pilot sequence is then transmitted in the pilot part of the transmission frame, namely, the first $n_p$ time instances. Note that each column of $\mathbf{A}$ is normalized to have an Euclidean norm of $\sqrt{n_p P_p}$ where $P_p$ is the average symbol power of the pilot part. The received signal in the pilot part at the $m$-th AP can be written as

\begin{equation}
    \mathbf{Y}_{p,m} = \sum\limits_{i =1 }^{K_a} \mathbf{a}_i \mathbf{g}_{i,m} + \mathbf{Z}_{p,m},
\end{equation}

\noindent where $\mathbf{Y}_{p,m} \in\mathbb{C}^{n_p \times M_r}$, $\mathbf{a}_i $ is the pilot sequence of the $i$-th user, and $\mathbf{Z}_{p,m}$ is the matrix of first $n_p$ columns of $\mathbf{Z}_{m}$.

The rest $B_c = B -B_p$ bits are appended by $r$ cyclic redundancy check (CRC) bits, encoded by a $(n_c, B_c + r)$ polar code, and then modulated by quadrature phase shift keying (QPSK) to obtain the symbol sequence with length $n_d = n_c /2$, where $n_c$ is the code length. The encoded and modulated sequence is then distributed to the data part of the frame via a transmission pattern determined by the first $B_p$ message bits. The transmission patterns are selected independently from the transmission pattern matrix $\mathbf{P} \in \mathbb{R}^{(n-n_p) \times N}$ with $n_d$ non-zeros in each column, denoting where the codeword elements will be placed (active indices). The received signal in the data part at the $m$-th AP can be written as



\begin{equation}
    \mathbf{Y}_{d,m} = \sum\limits_{i =1 }^{K_a} {s}_d(\mathbf{c}_i) \mathbf{g}_{i,m} + \mathbf{Z}_{d,m},
\end{equation}

\noindent where $\mathbf{Y}_{d,m} \in\mathbb{C}^{(n - n_p) \times M_r}$, $s_d(.)$ is a mapper distributing the modulated polar codeword of the $i$-th user $\mathbf{c}_i \in\mathbb{C}^{n_d \times 1} $ to the data part, and $\mathbf{Z}_{d,m}$ is the matrix consisting of the last $n - n_p$ rows of $\mathbf{Z}_{m}$. The elements of $\mathbf{c}_i$ are QPSK symbols in $\{ \sqrt{P_d/2} (\pm 1 \pm j)\}$ where $P_d$ is the average symbol power of the data part.


\subsection{Receiver Operation at AP}

At each AP, the received signal is processed to recover the selected pilot sequences, estimate the channel vectors, and estimate the transmitted data symbols using the channel vector estimates. The symbol estimates are then passed to the CPU via fronthaul, where the final decisions are made. To keep the system scalable, only $K_m$ users are recovered at each AP. In the following, we explain active pilot detection, channel estimation, and symbol estimation processes in detail.

First, we jointly estimate the active pilot sequences and the corresponding channel coefficients. Since $K_a \ll N$, this is essentially a compressed sensing problem with $\mathbf{A}$ being the measurement matrix. We apply the greedy iterative OMP algorithm \cite{omp} to solve this problem. In OMP, the element of the measurement matrix that has the highest correlation with the received signal is found and its effect is subtracted by using its projection onto the signal space at each iteration. Hence, in the first step, we calculate the correlation between the candidate pilots and the received signal at the $m$-th AP as


\begin{equation}
    \mathbf{R}_m = \mathbf{A}^H \mathbf{Y}_{p,m}^{(k)},
    \label{eqomp}
\end{equation}

\noindent where $\mathbf{R}_m \in\mathbb{C}^{N \times M_r}$, $\mathbf{Y}_{p,m}^{(k)}$ is the residual pilot signal at the $k$-th OMP iteration, and $\mathbf{Y}_{p,m}^{(1)} = \mathbf{Y}_{p,m}$. The decision metric of each pilot is obtained by calculating the Euclidean metric of the corresponding row, and the pilot index with the best decision metric is added to the output list $\mathcal{\hat{I}}_m$. Then, the effect of the detected pilots is subtracted as

\begin{equation}
\vspace{-2mm}
 \resizebox{.90\hsize}{!}{$\mathbf{Y}_{p,m}^{(k + 1)} = \mathbf{Y}_{p,m} - \mathbf{A}_{\hat{{I}}_m^{(k)}} \left(\mathbf{A}_{\hat{{I}}_m^{(k)}}^H\mathbf{A}_{\hat{{I}}_m^{(k)}} + N_0\mathbf{I}_{\abs{{\hat{I}}_m^{(k)}}} \right)^{-1} \mathbf{A}_{\hat{{I}}_m^{(k)}}^H \mathbf{Y}_{p,m}$}, 
    \label{eqsubt}
\end{equation}


\noindent where $\mathbf{A}_{\hat{{I}}_m^{(k)}}$ is the set of the columns of $\mathbf{A}$ specified by ${\hat{{I}}_m^{(k)}}$, $\hat{I}_m^{(k)}$ is the set of detected pilot indices until the $k$-th iteration, and $\mathbf{I}$ denotes the identity matrix. Note that we employ an LMMSE criterion in the subtraction step of OMP. Also, since the transmission patterns in the data part are also determined by the first $B_p$ bits, they are recovered in this step as well.
After $K_m$ iterations, the estimated channel vectors become

\begin{equation}
        \mathbf{\hat{G}}_m = \left(\mathbf{\hat{A}}_m^H\mathbf{\hat{A}}_m + N_0\mathbf{I}_{{K_m}} \right)^{-1} \mathbf{\hat{A}}_m^H \mathbf{Y}_{p,m}^{(j)},
    \label{eqest}
\end{equation}

\noindent where $\mathbf{\hat{G}}_m \in \mathbb{C}^{K_m \times M_r}, $ $\mathbf{\hat{A}}_m \in\mathbb{C}^{n_p \times K_m}$ is the detected pilot set at the $m$-th AP, and $\mathbf{Y}_{p,m}^{(j)}$ is the residual pilot signal at the $m$-th AP in the $j$-th decoding iteration.

Given the channel vector estimates, an LMMSE solution is employed to estimate the transmitted symbols in the data part. To do that, first, the LMMSE matrix can be calculated as

\begin{equation}
    \mathbf{W}_m = \left(\mathbf{\hat{G}}_m^H \mathbf{\hat{G}}_m + \frac{\sigma^2}{P_d} \mathbf{I}_{M_r} \right)^{-1} \mathbf{\hat{G}}_m^H.
    \label{eqsymbolest}
\end{equation}

Then, the symbol estimates are obtained by taking the symbols at the active indices at each row of 


\begin{equation}
    \hat{\mathbf{C}}'_m = \mathbf{Y}_{d,m}^{(j)} \mathbf{W}_m,
\end{equation}

\noindent where $\hat{\mathbf{C}}'_m \in \mathbb{C}^{(n - n_p) \times K_m}$, $\mathbf{Y}_{d,m}^{(j)}$ is the data part of the received signal at the $m$-th AP in the $j$-th decoding iteration.

The estimated symbols are passed to the CPU without estimated large-scale fading channel coefficients; namely, we consider \textit{Level 2} cooperation between the APs as in \cite{cefura}. The active pilot indices are also passed since they are required to combine the symbol estimates belonging to the same user coming from different APs.

\subsection{Symbol Combining and Decoding at CPU}

At the CPU, the symbol estimates from different APs are combined, and the corresponding log-likelihood ratio (LLR) values are extracted. The LLR values are fed to a single-user polar decoder employing successive cancellation list decoding (SCLD), and the decoded sequence is added to the output list if the CRC check is satisfied. The symbol estimates of the $i$-th user can be obtained by averaging over the following sum

\begin{equation}
    \mathbf{\hat{c}}_i =  \sum\limits_{m =1}^{M} \mathbbm{1}_{\{\text{ind}(\mathbf{\hat{a}}_i) \in \mathcal{\hat{I}}_m\}} \hat{\mathbf{C}}_m \left[:,f_m(\text{ind}(\mathbf{\hat{a}}_i)) \right],
\end{equation}

\noindent where $\hat{\mathbf{C}}_m \in \mathbb{C}^{n_d \times K_m}$ is the set of the estimated symbols, $\text{ind}(\mathbf{\hat{a}}_i)$ is the index of the estimated pilot sequence of the $i$-th user, and $f_m (.)$ is a function mapping the index of the estimated pilot to the corresponding column of $\hat{\mathbf{C}}_m$. Note that each pilot index estimate is matched to a user ignoring the collisions, as the number of pilot sequences can be set as to make the collision probability small. 

\subsection{Successive Interference Cancellation}

After the operation at the CPU is completed, the successfully recovered message bits are passed back to the APs with their active pilot indices via the backhaul, and re-encoded and modulated prior to SIC. SIC is a crucial step in the URA schemes to cope with high multiuser interference and realize low error rates. Without loss of generality, the SIC at the $m$-th AP is performed as follows

\begin{equation}
        \mathbf{Y}_m^{(j + 1)} = \mathbf{Y}_m^{(j)} - \sum\limits_{i =1}^{\abs{{\mathcal{\hat{D}}^{(j)}}}} \mathbbm{1}_{\{\text{ind}(\mathbf{\hat{a}}_i) \in \mathcal{\hat{I}}_m\}} \mathbf{\hat{x}}_{{\mathcal{\hat{D}}}^{(j,i)}}  \mathbf{\hat{g}}_{i,m},
        \label{eqsic}
\end{equation}

\noindent where $\mathbf{Y}_m^{(j)}$ is the residual of the received signal at the $m$-th AP,  ${\hat{\mathcal{D}}^{(j)}}$ is the set of successfully decoded messages, $\mathbf{\hat{x}}_{{\mathcal{\hat{D}}}^{(j,i)}}$ is the re-constructed transmitted signal of the $i$-th user at the $j$-th decoding iteration, and $\mathbf{\hat{g}}_{i,m}$ is its estimated channel vector. The decoding iterations continue until no new user message can be decoded successfully in the current iteration or $n_{\text{dec}}$ iterations are reached. The decoding procedure of the proposed scheme is illustrated in Fig. \ref{figdecoder}.

\subsection{Complexity Analysis}

In this subsection, we provide a computational complexity analysis of the proposed scheme in terms of the number of multiplications in decoding operations. The complexity of joint pilot and channel estimation is dominated by the matrix multiplication in (\ref{eqomp}) with a complexity of $\mathcal{O} (Nn_pM_r)$, where $\mathcal{O}(.)$ denotes the standard big-O notation. The complexity of symbol estimation is $\mathcal{O} (n_dM_rK_m)$, and SIC has a complexity of $\mathcal{O} ((n_d + n_p)K_mM_r)$, that are determined by the matrix multiplications in (\ref{eqsymbolest}) and (\ref{eqsic}), respectively.



\section{Numerical Results} \label{results}
 
In this section, we evaluate the performance of the proposed scheme through Monte Carlo simulations. We take $n = 3200$ and $B = 100$. We employ Gaussian sequences as pilots where each element of a pilot is a zero-mean standard normal random variable. We utilize 5G polar codes with a code length of 1024, set the CRC length to 16, and the list size of the polar decoder to 8. We assume that the APs are distributed to a $D \times D$ area according to a Binomial Point Process where $D = 550  \hspace{1mm} m$.

For the path loss, we utilize the urban micro-cell propagation model in \cite{stand} with a center frequency of 2 GHz similar to \cite{cefura, cellf1}. Hence, the large-scale fading coefficient $\beta_{i,m}$ can be written as

\begin{equation}
    \beta_{i,m} [dB] = -30.5 - 36.7\hspace{1mm}\text{log}_{10}d_{im} + F_{im},
\end{equation}

\noindent where $d_{im}$ is the distance between the $i$-th user and the $m$-th AP, and $F_{im}$ is the shadow fading. We utilize the same shadow fading model with \cite{cefura}, namely, $F_{im} \sim \mathcal{CN} (0,16)$ and it is correlated from an AP to different users as \cite{stand}


                \begin{flalign}
                \mathbb{E} [F_{im}F_{kj}] =
            \begin{cases}
                           16 \times 2^{-d'_{ik}/9} & \text{if $m=j$} \\ 
                             0, & \text{if $m \neq j$}
                        \end{cases} 
                \end{flalign}

 \noindent where $d'_{ik}$ is the distance between the $i$-th and $k$-th user. The small-scale fading coefficients are generated by assuming a uniform linear array (ULA) at the APs with half-wavelength antenna spacing and using the spatial correlation matrix in (2.23) in \cite{massmimo}. (See Sections 2.2 and 2.6 in \cite{massmimo} for further details on correlated fading).                

\begin{figure} 
    \centering
    \includegraphics[width=1\linewidth]{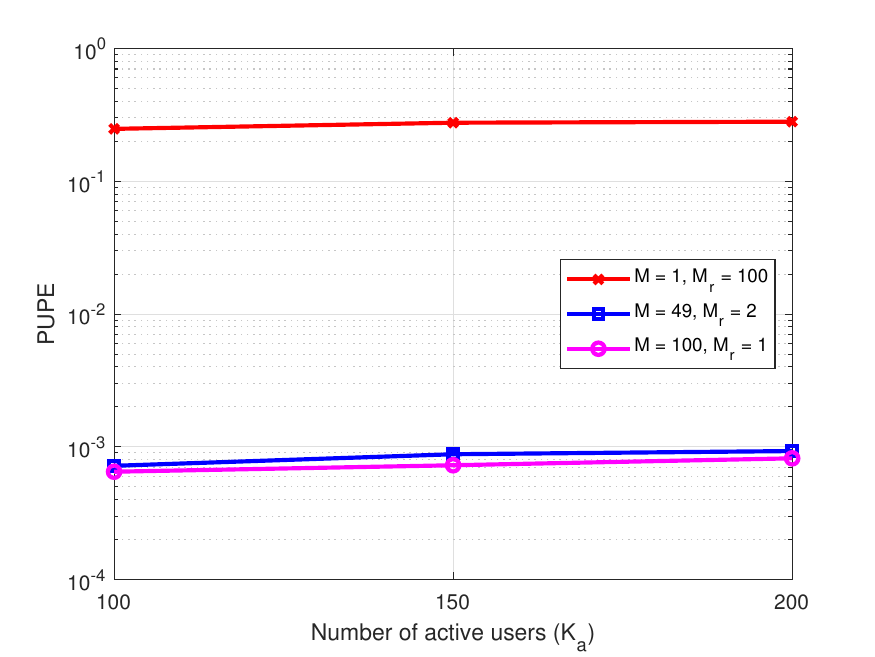}
    \caption{Comparison of PUPE versus the number of active users for different AP configurations.}
    \label{figdist}
\end{figure}

We assess the effect of distributing the receive antennas in Fig. \ref{figdist} for $K_m = 7$ by evaluating the PUPE of the proposed scheme for $M = 1$ and $M_r = 100$, i.e., the centralized scenario and compare it with that of the two distributed scenarios where $M = 49$ and $M_r = 2$ and $M = 100$ and $M_r = 1$. We set the average symbol power to 10 mW and $\sigma^2 = -84$  dBm. The results in Fig. \ref{figdist} show that distributing the antennas to the area through the APs significantly improves the performance, namely, the PUPE decreases by more than two orders of magnitude as the system becomes more robust to large-scale fading with the distribution of the APs to the geographic area.

\begin{figure} 
    \centering
    \includegraphics[width=1\linewidth]{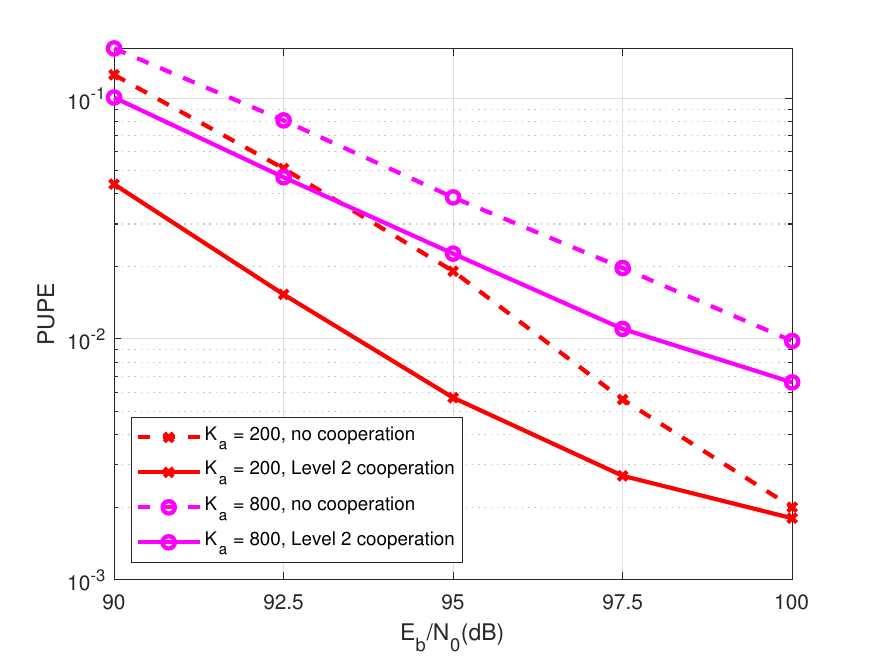}
    \caption{PUPE comparison of different cooperation levels.}
    \label{figcoop}
\end{figure}

We also investigate the effect of cooperation between the APs in Fig. \ref{figcoop} where we take $K_a = 200, 800$, $M = 100$ and $M_r = 1$ and compare the proposed scheme with the scenario that there is no cooperation between the APs, namely, each AP tries to recover the message of $K_m$ users without passing the symbol estimates to CPU. As expected, the AP cooperation improves the performance considerably even if only the symbol estimates are passed to the CPU. 

\begin{figure} 
    \centering
    \includegraphics[width=1\linewidth]{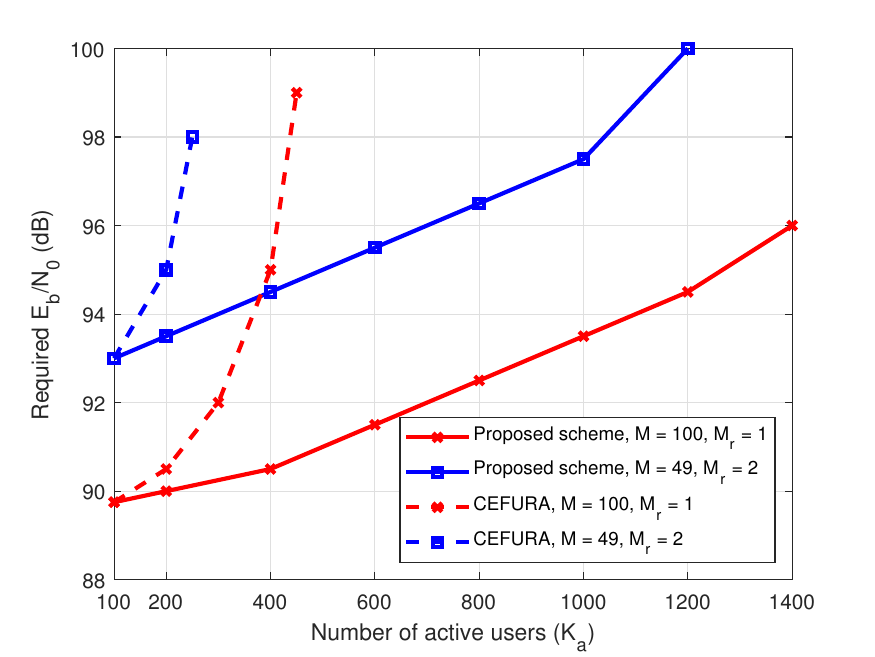}
    \caption{Comparison of the required $E_b/N_0$ versus the number of active users for PUPE $P_e \leq 0.05.$}
    \label{figreq}
\end{figure}

Finally, we evaluate the energy efficiency of our proposed scheme by calculating the required $E_b/N_0$ for a target PUPE of 0.05 in a wide range of active user load and compare it with the energy efficiency of the scheme in \cite{cefura} (CEFURA) for $K_m = 10$. The required transmit energy-per-bit of the system can be calculated as

\begin{equation}
    \frac{E_b}{N_0} = \frac{n_pP_p + n_dP_d}{B\sigma^2}.
\end{equation}


Note that the receive signal-to-noise ratio is much lower compared to transmit energy-per-bit due to the severe path-loss and shadowing. The results in Fig. \ref{figreq} demonstrate that the proposed scheme outperforms CEFURA for a maximum of 4.5 dB for $K_a \leq 400$ for $M = 100$ and $M_r = 1$, and 1.5 dB for $K_a \leq 200$ for $M = 49$ and $M_r = 2$. Moreover, it can support up to 1400 active users, depending on the configuration.

\vspace{-2mm}
\section{Conclusions} \label{conclusion}

We investigate unsourced random access in cell-free massive MIMO systems and adapt an ODMA-type solution combining OMP-based joint pilot and channel estimation and symbol estimation by MMSE at the APs. Symbol estimates are combined at the CPU assuming \textit{Level 2} cooperation, and the recovered bits are fed back to APs for SIC. Numerical results show that the cell-free structure considerably improves the system performance, and the proposed scheme offers excellent performance. In this work, we study one cooperation level between the APs. A deeper investigation of the cooperation levels and a theoretical analysis of the system performance can be interesting future directions.

\section*{Acknowledgement}

Mert Ozates' and Eduard Jorswieck's work is supported by European Commission’s Horizon Europe, Smart Networks and Services Joint Undertaking, research and innovation program under grant agreement number 101139282, 6G-SENSES project. Mohammad Kazemi’s work is funded by UKRI under the UK government’s Horizon Europe funding guarantee [grant number 101103430].




\end{document}